\documentclass[aps,prd,10pt,reprint,nofootinbib,floatfix]{revtex4-1}

\usepackage{graphicx}
\usepackage{dcolumn}
\usepackage{amssymb}
\usepackage{amsmath}
\usepackage{amsfonts}
\usepackage{amsbsy}
\usepackage{color}
\usepackage{rotating}
\usepackage{lipsum}
\usepackage[english]{babel}
\usepackage{soul}
\usepackage{multirow}
\usepackage{comment}
\usepackage{overpic}
\usepackage{silence}
\usepackage[english]{babel}
\usepackage[utf8]{inputenc}
\usepackage[dvipsnames]{xcolor}

\usepackage{soul}
\usepackage{cancel}
\usepackage{verbatim} 

\newcommand{\be}{\begin{equation}}
\newcommand{\ee}{\end{equation}}
\newcommand{\bea}{\begin{eqnarray}}
\newcommand{\eea}{\end{eqnarray}}

\usepackage[colorlinks]{hyperref}
\AtBeginDocument{\hypersetup{
        linkcolor=Blue,
        allcolors=Blue
}}
\usepackage{orcidlink}

\newcommand*{\rom}[1]{\expandafter\@slowromancap\romannumeral #1@}
\makeatother


\begin{document}

\title{Is Chevallier-Polarski-Linder dark energy a mirage?}

\author{Mikel Artola\orcidlink{0009-0007-9068-1995}} 
\email[Contact author: ]{mikel.artola@ehu.eus}
\author{Ismael Ayuso\orcidlink{0000-0002-0606-764X}} 
\email[Contact author: ]{ismael.ayuso@ehu.eus}
\author{Ruth Lazkoz\orcidlink{0000-0001-5536-3130}}
\email[Contact author: ]{ruth.lazkoz@ehu.eus}
\affiliation{Department of Physics, \href{https://ror.org/000xsnr85}{University of the Basque Country EHU}, 48940 Leioa, Spain}
\affiliation{EHU Quantum Center, \href{https://ror.org/000xsnr85}{University of the Basque Country EHU}, Leioa, 48940 Leioa, Spain}
\author{Vincenzo Salzano\orcidlink{0000-0002-4905-1541}}
\email[Contact author: ]{vincenzo.salzano@usz.edu.pl}
\affiliation{Institute of Physics, \href{https://ror.org/05vmz5070}{University of Szczecin}, Wielkopolska 15, 70-451 Szczecin, Poland}

\begin{abstract}
Recent observations from the Dark Energy Spectroscopic Instrument (DESI) raise doubts about the standard cosmological model, $\Lambda$CDM, suggesting a preference for an inherently dynamical dark energy component. The Chevallier-Polarski-Linder (CPL) parametrization---a widely used two-parameter model for the dark energy equation of state---displays marked early-time phantom behavior and a recent crossing of the phantom divide. These features suggest the convenience to check observationally the robustness of such evolution. To address this, we design two alternative families of two-parameter dark energy parametrizations which remain close to the original CPL but aim to soften its phantom character. Specifically, these models reproduce CPL-like behavior at low redshift but mitigate early phantom behavior through the use of smooth sigmoid transitions, yielding a more gradual evolution. By combining recent DESI data with constraints from the cosmic microwave background and Type Ia supernovae, we assess the viability of these models. Our analysis shows that CPL remains a strong and competitive parametrization, with the proposed alternatives only marginally favored or disfavored. We conclude that current observational data lack the statistical precision to decisively distinguish between CPL and similarly constructed parametrizations across the redshift range probed by late-time observables.

\vspace{.7em}
\noindent{\small DOI: \href{https://doi.org/10.1103/rz3s-zz61}{10.1103/rz3s-zz61}}

\end{abstract}

\maketitle

\noindent

\vspace{1em}

\section{Introduction}

Dark energy (DE)~\cite{Peebles:2002gy}, the dominant component of the current universe’s energy content, might reflect more than a transient stage in our current cosmological model. It could instead point to a fundamental aspect of the Universe itself. Its physical origin remains cloaked in theoretical intrigue. Broadly, it is framed either as an effective fluid called DE---perhaps a scalar field~\cite{Copeland:2006wr}---or as an arcane reinterpretation of gravity~\cite{Joyce:2016vqv}. From a dynamical standpoint, the distinction is twofold; either time evolution is manifest, or it is entirely absent. Nevertheless, it is convenient to clarify that time evolution may refer to two different aspects. On the one hand, in the fluid picture (whether effective or not) we may be open to consider a scale factor dependence in the fraction between the pressure and the energy density of the fluid, the so-called $w(a)$. This is the customary definition of evolving DE, and some of the earliest proposals in this direction are introduced in~\cite{Wang:2004xz}. On the other hand, the mentioned ratio may be constant and still correspond to a time-changing energy density due to the expansion of the Universe, unless the ratio takes the very specific case leading to a cosmological constant ($\Lambda$) for which $w_\Lambda(a) \equiv -1$. Precisely, the latter has been historically the most consensual case. But the mounting evidence on a redshift dependent $w$ of DE from the DESI Collaboration releases~\cite{DESI:2024mwx,DESI:2025zgx} has shaken the ground on which the $\Lambda$ and cold dark matter ($\Lambda$CDM) model stood firmly. Even so, such nondynamical interpretation of DE continues to be backed up by reliable surveys such as ACT~\cite{ACT:2025fju,ACT:2025tim}, so the debate continues.

In any case, for any eventual preferred realization of the behavior of DE, its dominance of the matter-energy budget makes it play a central role in the late-time dynamics of the Universe. Precisely, the physics describing this epoch took a fantastic turn when such mysterious agent responsible for repulsive gravity was discovered~\cite{Padmanabhan:2004av}. This revelation was prompted by the unexpected observation that the expansion of the Universe is accelerating, contrary to the deceleration expected from matter and relativistic species alone. The acceleration arises from both the relative abundance and the peculiar properties of DE, which together overcome the gravitational pull of other components.

One of the key challenges posed by DE is that it cannot be observed directly. Instead, its existence and effects are inferred through its influence on the expansion history of the universe [as revealed by Type Ia supernovae (SNeIa), baryon acoustic oscillations (BAO) and the cosmic microwave background (CMB)] and on the growth of cosmic structure (as traced by weak gravitational lensing and galaxy clustering). Besides, and very broadly, any specific DE paradigm must be compliant with independent age estimates of the Universe, such as those derived from globular clusters~\cite{Valcin:2020vav,Valcin:2021jcg,Valcin:2025luu}.

In addition to the considerations above, one of the long standing problems is modeling its early time behavior, where its effects are less obvious. The ubiquitous Chevallier-Polarski-Linder (CPL) parametrization~\cite{Chevallier:2000qy, Linder:2002et} of the equation of state (EOS), $w(a) = w_0 + w_a(1 - a)$, offers a minimalistic interpolation between the current value and that in the faraway past. Despite its simplicity, it allows for strong constraints (particularly on $w_0$), performs in a statistically favored way compared to $\Lambda$, and captures the essence of the dynamical nature of DE without requiring elaborate theoretical constructions. In fact, this model can be interpreted as the first-order Taylor expansion of $w(a)$ around the present time ($a = 1$) of a more complex model.

Nevertheless, the early time phantom nature hinted by the recent constraints coming from joint probes pivoting around DESI BAO justifies a closer look at the implications of this parametrization. This need is reinforced by the lack of a strong theoretical motivation supporting this unconventionally strong negativity of the pressure of DE at early times. That is accompanied with the observational conclusion that a recent phantom divide crossing has occurred as well. This prompts two questions. How pronounced must the phantom behavior be at high redshift? Is the phantom divide crossing an observationally favored feature?

We address these questions through alternative parametrizations which, despite being very close to CPL, contribute to the debate of whether this golden standard offers a decisively good depiction of high redshift DE, or rather it is a misleading depiction of the true evolutionary profile of DE in the past. In this vein, we demand our new parametrizations to resemble a cosmological constant more closely than the CPL model at asymptotic high redshifts, thereby hopefully mitigating somehow the large absolute values of $w$ that often result from CPL extrapolations, while ensuring they coincide with this evolutionary model in the vicinity of the present value of the scale factor, $a = 1$.

We organize our paper as follows. In Sec.~\ref{sec:parametrizations} we deliver a brief general discussion on DE parametrizations from the fluid description. Then, we present the main features of our parametrizations, which are based on the use of sigmoid functions, and discuss why they are suitable candidates to explore why CPL performs so well  on the one hand and to hint at shortcomings of constant $w$ scenarios on the other hand. In Sec.~\ref{sec:data} we present the data used and the followed methodology, and Sec.~\ref{sec:results} is devoted to the presentation and discussion of our findings. We offer our final insights in Sec.~\ref{sec:conclusions}.

\section{Why and how to specify \texorpdfstring{$w(a)$}{w(a)}} \label{sec:parametrizations}

Throughout this work we will follow the prescription of a set of cosmological fluids over a homogeneous and isotropic Friedman-Lemaître-Robertson-Walker background. Setting $c = 1$ hereafter, its line element for a spatially flat universe, motivated by the strong constraints on the spatial curvature~\cite{Planck:2019nip,Planck:2018vyg}, is
\begin{equation}
    \mathrm{d}s^2 =
    -\mathrm{d}t^2 + {a(t)}^2 \left( \mathrm{d}x^2 + \mathrm{d}y^2 + \mathrm{d}z^2 \right),
\end{equation}
where $(x,y,z)$ are the spatial coordinates, $t$ is the cosmological time and $a(t)$ is the scale factor. If the universe is populated with various species, each contributing to total energy density $\rho$ and pressure $p$, the Friedmann equations that govern the scale factor's evolution are
\begin{align}
    3H^2 &= 8\pi G \rho, \label{eq:Friedmann}\\
    \dot{H} &= -4\pi G \left( \rho + p \right). \label{eq:Raychaudhuri}
\end{align}
Here, $H \equiv \dot{a}/a$ is the Hubble factor, where a dot denotes derivatives with respect to $t$. When the different species do not interact with each other, their individual pressures $p_i$ and energy densities $\rho_i$ obey the conservation equation,
\begin{equation}\label{eq:continuity_equation}
    \dot{\rho}_i + 3H(\rho_i + p_i) = 0,
\end{equation}
where $\rho \equiv \sum_i\rho_i$ and $p \equiv \sum_i p_i$. While the barotropic EOS $p = w\rho$ is now standard in cosmology, and the $w_i \equiv p_i/\rho_i$ is commonly used to characterize individual components, it appears that this specific notation was first introduced in~\cite{Turner:1997npq}.

Here we assume that the other companions of DE in the cosmic budget are baryonic and cold dark matter and radiation, so their pressure-to-energy-density ratios have fixed known values: $w_\mathrm{m} = 0$ for the matter content and $w_\mathrm{r} = 1/3$ for radiation. In other words, assuming that the scale factor just now satisfies $a(t_0) \equiv a_0 = 1$ and solving Eq.~\eqref{eq:continuity_equation}, matter and radiation evolve as
\begin{equation}
    \rho_\mathrm{m}(a) = \rho_{\mathrm{m},0} a^{-3}
    ,\qquad
    \rho_\mathrm{r}(a) = \rho_{\mathrm{r},0} a^{-4},
\end{equation}
where $\rho_{i,0}$ stands for the current value of the energy density of the $i$th species. Therefore, we can lighten the notation to let $w(a)$ simply stand for the EOS of DE. The energy density of DE is thus uniquely characterized by the parametrization of $w(a)$,
\begin{equation}\label{eq:DE_energy_density}
    \rho_\mathrm{DE}(a) =
    \rho_\mathrm{DE,0}\exp\left( 3 \int_a^1 \frac{1 + w(a^\prime)}{a^\prime}\, \mathrm{d}a^\prime \right).
\end{equation}
It is clear from Eq.~\eqref{eq:DE_energy_density} that the evolutionary features of DE, and therefore those of the Hubble function through Eq.~\eqref{eq:Friedmann}, depend crucially on the functional form of $w(a)$.

DE parametrizations in general are a standard route to give coherence to cosmological data that, in their original state, are scattered and unpatterned. In theory, this formulation is well defined and typically offers useful and not too technical information, in particular as compared to nonparametric models, which lay beyond the focus of this work despite their undeniable complementary interest.

Here we present and test a new class of DE parametrizations stemming from a simple idea. Much of the literature is based exactly on the famous CPL parametrization,
\begin{equation}
    w(a) =
    w_0 + w_a(1 - a),
\end{equation}
which can be understood as the first order of the Taylor series of an arbitrary EOS and therefore $w_0 \equiv w\vert_{a = 1}$. Nonetheless, many other explorations have considered scenarios displaying a behavior close to CPL for $a \sim 1$ while keeping just two DE parameters (see~\cite{Jassal:2004ej,Barboza:2008rh,Efstathiou:1999tm,Dimakis:2016mip}). 
Our proposal seeks to extend this similarity a little further beyond the strict vicinity of $a = 1$, and at the same time give $w(a)$ the possibility to perform nonlinearly in the $a \ll 1$ regime. Specifically, given the fact that current surveys point towards a strongly negative pressure of DE in the early Universe when using the CPL parametrization, we aim to asses whether the high-redshift behavior of DE could remain closer to a cosmological constant.

Before proceeding, let us remember what $w_a$ seems to tell us about the nature of DE in the CPL parametrization. Following the standard definition of the scale factor in terms of the redshift $z$, $a \equiv (1 + z)^{-1}$, and as discussed in~\cite{Linder:2002et}, it is argued that the quantity,
\begin{equation}
    \left. \frac{\mathrm{d} w}{\mathrm{d} \ln(1 + z)} \right|_{z=1} = \frac{w_a}{2},
\end{equation}
may serve as a useful indicator of the time variation of the EOS. 

Turning back to our proposal, we will resort to odd, monotonic and smooth sigmoid functions $\sigma(x)$ centered at the origin such that the pressure-to-energy-density ratio is parametrized as
\begin{equation}
    w(a) = w_0 + w_a\sigma(1-a).
\end{equation}
In order to ensure that $w(a)$ resembles that of CPL at small redshifts, that is, $\sigma_\mathrm{CPL}(1 - a) = 1 - a$, we demand these to satisfy,
\begin{equation}
    \sigma(1 - a)  \simeq (1 - a) + \mathcal{O}\left( (1 - a)^2 \right).
\end{equation}
As  mentioned earlier, we are interested in functions with early-time CPL deviations which render them somewhat closer to  a cosmological constant behavior. Precisely, given that recent surveys point towards phantom DE in the past, with $w_a < 0$ (and $w_0 < -1/3$, as it corresponds to a currently accelerated universe), this can be accomplished by demanding $\sigma(1 - a) < 1 - a$ in the range $a \in [0, 1)$. Clearly, these conditions can be fulfilled by a large variety of functions. 

To further control their behavior, we will restrict the analysis to suitably chosen sigmoid functions which do not present any inflection point in the range $a \in [0,1)$; that way, bounding the slope of $\sigma(1 - a)$ appropriately ensures $\sigma(1 - a) < 1 - a$. In order to do so, we target functions which satisfy these additional requirements for $a \in [0,1)$,
\begin{equation}\label{eq:slope_condition}
    \left| \frac{\mathrm{d}}{\mathrm{d}a} \sigma(1-a) \right| < 1;
    \qquad
    \frac{\mathrm{d}}{\mathrm{d}a} \sigma(1-a)< 0.
\end{equation}
The former is a very standard characteristic for this type of functions and softens their evolutionary features, whereas the latter is imposed for consistency with the slope of the CPL parametrization. These conditions ensure that we have an excellent late universe CPL mimicry with the possibility of exhibiting significantly discrepant early behavior. Indeed, assuming a setting with $w_0, w_a<0$ we can see that for identical values of those two parameters the following holds:
\begin{equation}
    w_\infty \equiv
    -|w_0 + w_a \sigma(1)| >
    -|w_0 + w_a|,
\end{equation}
where $w_\infty$ corresponds to the value of the EOS at infinitely large redshift $(a = 0)$. Hence, we seek for parametrizations that soften the phantomness of DE across the cosmological history and allow to test the robustness of the CPL parametrization. Presumably, parameter constraining pipelines equivalent to the DESI ones will not yield dramatically different $w_a$ values as our new parametrizations describe mild departures from the CPL paradigm.

Now we are readier to explore our parametrizations further. Many alternatives for our choice of the sigmoid function are available, but in this work we propose two families of parametrizations, which we will refer to as ``$\mathrm{Sqrt}n$'' and ``$n \mathrm{Tanh}$'',
\begin{align}
    \sigma_{\mathrm{Sqrt}n}(1-a) &= \frac{1-a}{\sqrt[2n]{1 + {(1-a)}^{2n}}}, \label{eq:Sqrtn} \\[5pt]
    \sigma_{n\mathrm{Tanh}}(1-a) &= n\tanh \left( \frac{1-a}{n} \right). \label{eq:nTanh} 
\end{align}
Here we consider $n \in \mathbb{Z}^+$, and the CPL parametrization is simply given by
\begin{equation}
    \sigma_\mathrm{CPL}(1-a) = 
    1 - a.
\end{equation}
These functions are constructed such that the first order in the Taylor expansion around $a = 1$ matches CPL's, $\sigma_{n\mathrm{Tanh}}(1-a) \simeq 1 - a$ and $\sigma_{\mathrm{Sqrt}n}(1-a) \simeq 1 - a$, ensuring that they follow the trend of the linear parametrization for most of the redshift interval of local measurements. Moreover, it is easy to see that their derivatives also satisfy the requirements of Eq.~\eqref{eq:slope_condition}, and therefore, for equal values of the EOS parameters $(w_0, w_a)$, these EOS will have a smaller absolute value of $w_\infty$ than CPL.

Notably, both parametrizations mimic CPL for a wider range of redshifts as $n$ increases, since in the limit $n \to \infty$ these tend exactly to CPL. Contrary to the linear parametrization, which is ill-defined in the asymptotic future ($a \to \infty$), the parametrizations proposed are bounded over all the cosmological history,
\begin{equation}
    \lim_{a \to \infty} \sigma_{\mathrm{Sqrt}n} = -1
    ;\qquad
    \lim_{a \to \infty} \sigma_{n\mathrm{Tanh}} = -n
    .
\end{equation}

\begin{figure*}[t]
    \centering
    \includegraphics{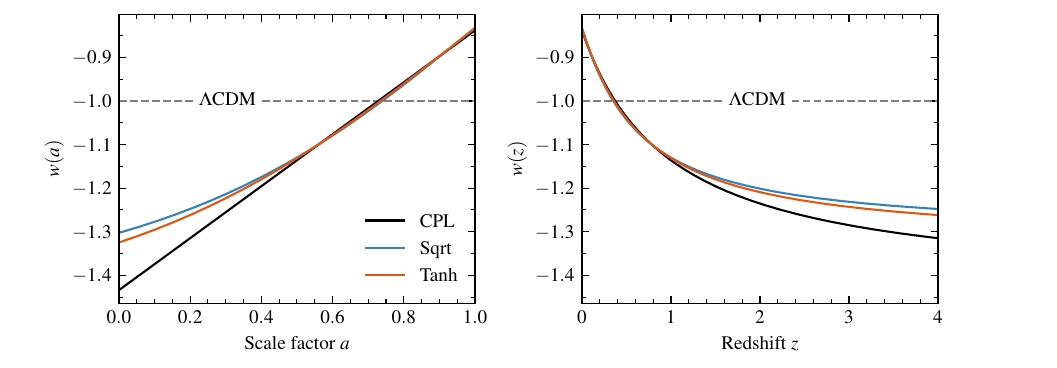}
    \vspace{-0.7em}
    \caption{Evolution of the DE EOS parameter $w$ for the CPL, Sqrt, and Tanh parametrizations, using the best-fit values of $(w_0, w_a)$ obtained from the combined dataset (CMB+BAO+SNeIa, see Sec.~\ref{sec:data} for details). The Sqrt and Tanh cases correspond to the representative choice $n = 1$ in Eqs.~\eqref{eq:Sqrtn} and~\eqref{eq:nTanh}. Left: Evolution across the full cosmic history as a function of the scale factor $a$. Right: Corresponding evolution in redshift space, where all models are nearly identical for $z \lesssim 1$.}%
    \label{fig:w_plot}
\end{figure*}

To illustrate the points discussed throughout this section, Fig.~\ref{fig:w_plot} displays the evolution of the EOS parameter for the CPL, Sqrt, and Tanh parametrizations. In the last two cases, we adopt the representative choice $n = 1$ in Eqs.~\eqref{eq:Sqrtn} and~\eqref{eq:nTanh}, as these are the ones which deviate furthest from CPL. The parameters $(w_0, w_a)$ are set to their best-fit values from the combination of CMB, BAO and SNeIa measurements; the data selection will be described in Sec.~\ref{sec:data}, and the corresponding constraints will be presented in Sec.~\ref{sec:results}. The left panel shows the evolution in terms of the scale factor, highlighting that Sqrt and Tanh lead to smaller deviations of $w_\infty$ from $-1$ than CPL. As noted before, this difference will decrease when considering larger values of $n$. The right panel shows the redshift dependence, where the three parametrizations are almost indistinguishable at low redshift $(z \lesssim 1)$. We shall discuss further quantitative results in Sec.~\ref{sec:results}.

Before closing this section, we note that related steplike extensions of the $n$Tanh form, introducing two additional parameters, were previously proposed in~\cite{Douspis:2008mxi, Linden:2008mf}. In general, three-parameter DE parametrizations continue to raise increasing interest~\cite{Alam:2025epg}, especially as error bars on the classic ones (that is, $w_0$ and $w_a$) tighten. However, the additional flexibility often comes at the cost of weakened inference; the conclusions may become less robust or require marginalization over the extra degrees of freedom~\cite{Linder:2007ka, Nesseris:2025lke, Malekjani:2025alf}, effectively reducing the parameter space back to two. This reinforces the motivation for adopting the simpler families proposed here.

\section{Data and methodology} \label{sec:data}

\subsection{Datasets}

To assess how well the parametrizations introduced in this study align with observations, we rely on data from BAO, SNeIa, and CMB.

\subsubsection{Baryon acoustic oscillations}

Regarding BAO, we use the data compilation from the DESI Data Release 2~\cite{DESI:2025zgx} (hereafter termed DESI). This set includes seven measurements of $D_\mathrm{M}/r_\mathrm{d}$, $D_\mathrm{H}/r_\mathrm{d}$ and $D_\mathrm{V}/r_\mathrm{d}$ for the range $0.295 \leq z_\mathrm{eff} \leq 2.33$. Collecting the cosmological parameters in a vector $\boldsymbol{p}$, $D_\mathrm{M}(z, \boldsymbol{p})$ is the transverse comoving distance,
\begin{equation}
    D_\mathrm{M}(z, \boldsymbol{p}) \equiv
    \int_0^z \frac{\mathrm{d}z^\prime}{H(z^\prime, \boldsymbol{p})},
\end{equation}
$D_\mathrm{H}(z, \boldsymbol{p}) \equiv 1/H(z, \boldsymbol{p})$ is the Hubble distance, $D_\mathrm{V}(z, \boldsymbol{p})$ is the angle-averaged comoving distance,
\begin{equation}
    D_\mathrm{V}(z, \boldsymbol{p}) \equiv
    \left( z D_\mathrm{M}(z, \boldsymbol{p})^2 D_\mathrm{H}(z, \boldsymbol{p}) \right)^{1/3},
\end{equation}
and $r_\mathrm{d} \equiv r_\mathrm{s}(z_\mathrm{d}, \boldsymbol{p})$ is the comoving sound horizon evaluated at the drag redshift $z_\mathrm{d}$. Instead of adopting the approximate formula for $r_\mathrm{d}$ as in~\cite{DESI:2025zgx}, we use its integral definition,
\begin{equation}\label{eq:BAO_sound_horizon}
    r_\mathrm{s}(z, \boldsymbol{p}) \equiv
    \int_z^\infty \frac{c_\mathrm{s}(z^\prime)\, \mathrm{d}z^\prime}{H(z^\prime, \boldsymbol{p})}.
\end{equation}
Here $c_\mathrm{s}(z)$ is the sound speed of the photon-baryon plasma before the recombination epoch,
\begin{equation}
    c_\mathrm{s}(z) =
    \frac{1}{\sqrt{3\left( 1 + \bar{R}_\mathrm{b}{(1 + z)}^{-1} \right)}},
\end{equation}
and $\bar{R}_\mathrm{b} = 31500 \Omega_\mathrm{b} h^2 (T_\mathrm{CMB}/2.7)^{-4}$ is the baryon-to-photon energy density ratio, with $T_\mathrm{CMB} = 2.726\ \mathrm{K}$, and $h \equiv H_0/(100\ \mathrm{km}\, \mathrm{s}^{-1}\, \mathrm{Mpc}^{-1})$. From this measurement of $T_\mathrm{CMB}$ and assuming the standard neutrino degrees of freedom ($N_\mathrm{eff} = 3.046$), we fix the fractional radiation density to the value $\Omega_\mathrm{r} = 4.18 \times 10^{-5} h^{-2}$. Following~\cite{Aizpuru:2021vhd}, we compute the redshift at the drag epoch using the following fitting formula:
\begin{equation}
    z_\mathrm{d} =
    \frac{1 + 428.169 \omega_\mathrm{b}^{0.256459} \omega_\mathrm{m}^{0.616388} + 925.56 \omega_\mathrm{m}^{0.751615}}{\omega_\mathrm{m}^{0.714129}},
\end{equation}
which is accurate up to $\sim\! 0.001\%$. Here we defined $\omega_\mathrm{b} = \Omega_\mathrm{b} h^2$ and $\omega_\mathrm{m} = \Omega_\mathrm{m} h^2$, with $\Omega_\mathrm{b}$ and $\Omega_\mathrm{m}$ the current values of the baryon and matter fractional energy densities, respectively.

In the absence of an external calibration of $r_\mathrm{d}$, BAO present a degeneracy between $H_0$ and $r_\mathrm{d}$, as it is readily seen from Eq.~\eqref{eq:BAO_sound_horizon}, being sensitive only to the combination $H_0 r_\mathrm{d}$. The CMB contains enough information to break this degeneracy. When performing the analysis without the CMB, we make use of a Gaussian prior on the fractional baryon density $\Omega_\mathrm{b}h^2$ coming from the big bang nucleosynthesis (BBN). More specifically, we adopt the value provided in~\cite{Schoneberg:2024ifp},
\begin{equation}\label{eq:BBN_prior}
    \Omega_\mathrm{b} h^2 =
    0.02218 \pm 0.00055.
\end{equation}

Then, we define the corresponding $\chi_\mathrm{BAO}^2$ as
\begin{equation}\label{eq:chi2_DESI}
    \chi_\mathrm{BAO}^2 =
    (\Delta \boldsymbol{\mathcal{D}}_\mathrm{BAO})^T \cdot \boldsymbol{C}_\mathrm{BAO}^{-1} \cdot \Delta \boldsymbol{\mathcal{D}}_\mathrm{BAO},
\end{equation}
where $\Delta \boldsymbol{\mathcal{D}}_\mathrm{BAO}$ is the difference between the theoretical and observational values of $(D_\mathrm{M}/r_\mathrm{d}, D_\mathrm{H}/r_\mathrm{d}, D_\mathrm{V}/r_\mathrm{d})$, and $\boldsymbol{C}_\mathrm{BAO}$ is the covariance matrix constructed using the errors and correlations from~\cite{DESI:2025zgx}. For the BGS tracer at the effective redshift $z_\mathrm{eff} = 0.295$ we only use the reported value of $D_\mathrm{V}/r_\mathrm{d}$, whereas for the rest of the tracers we use the transverse angular distance and Hubble distance pairs $(D_\mathrm{M}/r_\mathrm{d}, D_\mathrm{H}/r_\mathrm{d})$.

\subsubsection{Type Ia supernovae}

We make use of the Pantheon+ data~\cite{Scolnic:2021amr, Peterson:2021hel, Carr:2021lcj, Brout:2022vxf}. From this compilation, and following recent analyses from the DESI Collaboration~\cite{DESI:2024mwx,DESI:2025zgx,Peterson:2021hel}, we select light curves of SNeIa in the redshift range $0.01 < z < 2.26$ and compare them with the theoretical modulus, defined as
\begin{equation}
    \mu_\mathrm{theo} (z_\mathrm{hel}, z_\mathrm{HD}, \boldsymbol{p}) \equiv
    25 + 5\log_{10} D_\mathrm{L}(z_\mathrm{hel}, z_\mathrm{HD}, \boldsymbol{p}).
\end{equation}
In the latter, $D_\mathrm{L}$ is the luminosity distance (in Mpc),
\begin{equation}
    D_\mathrm{L}(z_\mathrm{hel}, z_\mathrm{HD}, \boldsymbol{p}) \equiv
    (1 + z_\mathrm{hel}) D_\mathrm{M}(z_\mathrm{HD},\boldsymbol{p}),
\end{equation}
$z_\mathrm{hel}$ is the heliocentric redshift, and $z_\mathrm{HD}$ is the Hubble diagram redshift obtained after correcting the cosmological redshift for the peculiar velocities of the galaxies and the solar system~\cite{Carr:2021lcj}. On the other hand, the observational distance modulus is $\mu_{\mathrm{obs},i} = m_{\mathrm{B},i}^0 - M$, where $m_{\mathrm{B},i}^0$ is the peak magnitude in the $i$th host and $M$ the fiducial absolute magnitude. Therefore, the difference between the theoretical and observed distance modulus is given by
\begin{equation}\label{eq:distance_modulus_difference}
    \Delta \mu_{\mathrm{SN},i} =
    m_{\mathrm{B},i}^0 - M - \mu_{\mathrm{theo},i}.
\end{equation}

Equation~\eqref{eq:distance_modulus_difference} presents a degeneracy between the fiducial absolute magnitude $M$ and the Hubble constant $H_0$. Excluding \textit{SH0ES}'s calibrated Cepheids due to their inconsistency with the CMB~\cite{Brout:2022vxf} demands the marginalization over the absolute magnitude. Following~\cite{Conley_2010}, the $\chi_\mathrm{SN}^2$ is formulated as
\begin{equation}\label{eq:chi2_SN}
    \chi_\mathrm{SN}^2 =
    a + \log \frac{d}{2\pi} - \frac{b^2}{d},
\end{equation}
where $a \equiv \Delta \boldsymbol{\mu}_\mathrm{SN}^T \cdot \boldsymbol{C}_\mathrm{SN}^{-1} \cdot \Delta \boldsymbol{\mu}_\mathrm{SN}$, $b \equiv \Delta \boldsymbol{\mu}_\mathrm{SN}^T \cdot \boldsymbol{C}_\mathrm{SN}^{-1} \cdot \boldsymbol{1}$, and $d ~\equiv~\boldsymbol{1}^T \cdot \boldsymbol{C}_\mathrm{SN}^{-1} \cdot \boldsymbol{1}$, and $\boldsymbol{C}_\mathrm{SN}$ is the covariance matrix (statistical plus systematic errors) reported by the Pantheon+ team; the $\Delta \boldsymbol{\mu}_\mathrm{SN}$ are those defined in Eq.~\eqref{eq:distance_modulus_difference}, but with the absolute magnitude $M$ omitted.

\subsubsection{Cosmic microwave background}

In order to include early-universe constraints, we use the compressed shift parameters introduced in~\cite{Wang:2007mza}. These encode the essential information of the full CMB power spectrum without requiring to compute the full perturbation theory, allowing for a much faster evaluation of the CMB likelihood. The shift parameters are defined as
\begin{align}
    R(\boldsymbol{p}) &\equiv
    \sqrt{\Omega_\mathrm{m} H_0^2}\, r(z_*, \boldsymbol{x}), \\
    l_a(\boldsymbol{p}) &\equiv \pi \frac{r(z_*, \boldsymbol{x})}{r_\mathrm{s}(z_*, \boldsymbol{x})},
\end{align}
where $r(z_*, \boldsymbol{x}) = D_\mathrm{M}(z_*, \boldsymbol{x})$ and $r_\mathrm{s}(z_*, \boldsymbol{x})$ is the transverse comoving distance and the sound horizon (respectively) evaluated at the recombination $z_*$, which we compute using the fitting formula given in~\cite{Aizpuru:2021vhd}
\begin{equation}
    z_* =
    \frac{391.672 \omega_\mathrm{m}^{-0.372296} + 937.422 \omega_\mathrm{b}^{-0.97966}}{\omega_\mathrm{m}^{-0.0192951} \omega_\mathrm{b}^{-0.93681}} + \omega_\mathrm{m}^{-0.731631}.
\end{equation}
The latter is accurate up to $\sim 0.0005\%$. 

We define the $\chi_\mathrm{CMB}^2$ as follows:
\begin{equation}\label{eq:chi2_CMB}
    \chi_\mathrm{CMB}^2 =
    (\Delta \boldsymbol{\mathcal{S}}_\mathrm{CMB})^T \cdot \boldsymbol{C}_\mathrm{CMB}^{-1} \cdot \Delta \boldsymbol{\mathcal{S}}_\mathrm{CMB},
\end{equation}
where $\Delta \boldsymbol{\mathcal{S}}_\mathrm{CMB}$ is the difference between the theoretical and observed values of $(R, l_a, \Omega_\mathrm{b} h^2)$, and  $\boldsymbol{C}_\mathrm{CMB}$ is the covariance reported in~\cite{Bansal:2025ipo}.%
\footnote{Though these values were obtained as a compression assuming a fiducial $\Lambda$CDM cosmology, we follow the assumption that the $\Lambda$CDM alternatives are in agreement with $\Lambda$CDM at early-times~\cite{Orchard:2024bve}. This is consistent with the results of $w_\infty$ in Table~\ref{tab:results}.}
The latter were computed based on PR$3$ \emph{Planck}~\cite{Planck:2019nip} and Data Release 6 of ACT~\cite{ACT:2023kun} datasets.

\subsection{Method and model comparison}

We perform Markov chain Monte Carlo (MCMC) runs with different combinations of the CMB, BAO and SNeIa data (denoted as C, B, and S, respectively) in order to minimize the total $\chi_\mathrm{tot}^2$, which is constructed as the sum of the $\chi^2$ of the datasets combined from Eqs.~\eqref{eq:chi2_DESI},~\eqref{eq:chi2_SN}, and~\eqref{eq:chi2_CMB}. More specifically, we test the parametrizations with the following dataset combinations, ordered from early- to late-time data; CMB$+$BAO$+$SNeIa (CBS), CMB$+$BAO (CB), CMB$+$SNeIa (CS), and BAO+SNeIa (BS). Note that recent studies report tensions between the selected surveys~\cite{Wang:2025bkk, Ye:2025ark}. In our analysis, we allow the cosmological parameters $\boldsymbol{p} = (\Omega_\mathrm{m}, H_0, \Omega_\mathrm{b}, w_0, w_a)$ to vary, imposing the uninformative flat priors summarized in Table~\ref{tab:priors} and the additional prior $w_\infty < 0$ in order to ensure that DE is a subdominant species in the past. Note that we do not impose a prior directly on $w_a$ as in~\cite{DESI:2024mwx, DESI:2025zgx} because the time evolution of $w(a)$ is modified by the sigmoid functions in Eqs.~\eqref{eq:Sqrtn} and~\eqref{eq:nTanh}. Instead, we opt for imposing the prior over the modified parameter $\tilde{w}_a \equiv w_a \sigma(1)$; for CPL, it reduces to the customary $w_a$. Conversely, for Sqrt$n$ and $n$Tanh it accounts for the effective rescaling of the asymptotic value of the sigmoid function at $a = 0$. After ensuring that the chains have truly converged after a final $100$K points run, we present the ``best-fit'' (strictly speaking, the median of the posterior) values of the parameters. The figures are produced using \texttt{Matplotlib}~\cite{Hunter:2007} and \texttt{GetDist}~\cite{Lewis:2019xzd}.

\begin{table}[t]
    \centering
    \caption{Prior distributions implemented; for shortness, we defined $\tilde{w}_a \equiv w_a \sigma_i(1)$ for the $i$th parametrization, which accounts for the modified evolution due to the addition of the sigmoid functions. Here $\mathcal{U}[x, y]$ represents a uniform distribution over the interval $[x, y]$.}%
    \label{tab:priors}
    \begin{ruledtabular}
    \begin{tabular}{clcc}
        & Parameter & Prior (flat) & \vspace{.1em} \\ \hline\rule{0pt}{1.1em}%
        & $\Omega_\mathrm{m}$ & $\mathcal{U}[0.2, 0.8]$ & \\[0.25em]
        & $H_0$ & $\mathcal{U}[20, 100]$ & \\[0.25em]
        & $\Omega_\mathrm{b} h^2$ & $\mathcal{U}[0.005, 0.1]$ & \\[0.25em]
        & $w_0$ & $\mathcal{U}[-3, 1]$ & \\[0.25em]
        & $\tilde{w}_a$ & $\mathcal{U}[-3, 2]$ & \\ 
    \end{tabular}
    \end{ruledtabular}
\end{table}

To rigorously evaluate the statistical performance of our models relative to the CPL parametrization, we compute the Bayes factor as a comparison tool, $\mathcal{B}_i^\mathrm{CPL}$, which we define in this case as the ratio between the Bayesian evidences $Z$ (that is, marginalized posteriors) of the CPL and the $i$th parametrization or model, $\mathcal{B}_i^\mathrm{CPL} \equiv Z_\mathrm{CPL}/Z_i$. These evidences are calculated using the nested sampling algorithm following~\cite{Mukherjee:2005wg}, and the comparison is carried out through Jeffrey's scale~\cite{Jeffreys61}; if $\ln \mathcal{B}_i^\mathrm{CPL} < 1$, the support in favor of CPL is considered weak or negligible; when $1 < \ln \mathcal{B}_i^\mathrm{CPL} < 2.5$, the evidence is considered substantial; for $2.5 < \ln \mathcal{B}_i^\mathrm{CPL} < 5$, the support for CPL is strong, and if $\ln \mathcal{B}_i^\mathrm{CPL} > 5$, the evidence in favor of CPL is regarded as decisive.

For ease of comparison with prior findings, we present the statistical significance of the CPL parametrization relative to model $i$ in terms of the number of standard deviations $\# \sigma$. This is achieved by calculating the relative difference $\Delta \chi^2 = \mathrm{min}( \chi_\mathrm{CPL}^2 ) - \min( \chi_i^2 )$, approximately described by a $\chi^2$ distribution with $\Delta f = f_\mathrm{CPL} - f_i$ degrees of freedom. Assuming both models fit the data equally well (null hypothesis), the probability of encountering $\Delta \chi^2$ is given by the $p$-value,
\begin{equation}
    p =
    1 - \mathcal{P}_{\chi^2}(\vert \Delta \chi^2|, \Delta f),
\end{equation}
where $\mathcal{P}_{\chi^2} (\alpha, f)$ is the cumulative distribution function for a $\chi^2$ distribution with $f$ degrees of freedom,
\begin{equation}
    \mathcal{P}_{\chi^2}(\alpha, f) =
    \frac{1}{2^{f/2} \Gamma(f/2)} \int_0^\alpha \mathrm{d}x\, x^{f/2-1} e^{-x/2},
\end{equation}
with $\Gamma(x)$ representing the gamma function. The $p$-value is then translated into the number of standard deviations, $\#\sigma$, using the inverse of standard normal distribution's cumulative distribution function,
\begin{equation}
    \sigma =
    \Phi^{-1}(1 - p/2)
    ;\quad
    \Phi(x) =
    \frac{1}{\sqrt{2\pi}} \int_{-\infty}^x \mathrm{d}y\, e^{-y^2/2}.
\end{equation}

\section{Results and discussion} \label{sec:results}

In this section we compare the Sqrt$n$ and $n$Tanh parametrizations with the CPL parametrization for the values $n \in \{1, \dots,4\}$. Table~\ref{tab:results} contains the main results of this work,%
\footnote{Our results can differ slightly from those reported in other works, e.g.,~\cite{DESI:2025zgx, Ishak:2025cay, Giare:2025pzu}. Such discrepancies may arise from using our own sampling pipelines, or from differences in the criteria followed to implement the data.}
with the best-fit values of the sampled cosmological parameters $(\Omega_\mathrm{m}, H_0, \Omega_\mathrm{b}, w_0, w_a)$, as well as the constraints on the derived parameter $w_\infty$ and the logarithm of the Bayes' factors $\ln \mathcal{B}_i^\mathrm{CPL}$.

We start by making a concise comparison of the CPL parametrization with the $\Lambda$CDM model and the $w_0$CDM model, for which the EOS is constant and equal to $w_0$. Then, we proceed to asses our novel parametrizations using CPL as a comparative standard. We first examine the statistical significance of our results by addressing the EOS behavior. Subsequently, we investigate the numerical outcomes for $(w_0, w_a)$ and the associated $w_\infty$, comparing across various dataset combinations as shown in Table~\ref{tab:results}. Lastly, we discuss the results for the remaining cosmological parameters $(\Omega_\mathrm{m}, H_0, \Omega_\mathrm{b})$ listed in Table~\ref{tab:results}.

\subsection{Evidence in favor of dynamical dark energy}

Table~\ref{tab:LCDM_w0CDM_CPL} contains the Bayes' factors $\ln \mathcal{B}_i^\mathrm{CPL}$ and the number of statistical deviations, $\#\sigma$, of $\Lambda$CDM and $w_0$CDM compared to CPL for the four dataset combinations. It can be readily seen that the single combination that does not include the DESI data, CS, shows no preference for dynamical DE, as the standard deviations are below $0.5\sigma$ and the logarithm of the Bayes' factors is slightly negative. In this case, both a cosmological constant and a constant EOS, whose best-fit results to be $w_0 = 0.975_{-0.029}^{+0.030}$, provide a good fit to the data.

This is no longer true when DESI data are considered. If the CMB distance priors are not included (BS combination) the tension with the cosmological constant increases to $1.7\sigma$, indicating a slight preference for evolving DE; nonetheless, a constant EOS $w_0 = -0.913 \pm 0.040$ performs equally well compared to the latter. The tension is at the highest when SNeIa are not considered (CB), increasing to $2.8\sigma$ for $\Lambda$CDM and $3\sigma$ for $w_0$CDM with $w_0 = -1.04_{-0.039}^{+0.037}$, and the logarithm of the Bayes' factors almost surpassing in this case the threshold where evidence can be regarded as decisive. When all datasets are combined the tension is minimally reduced to $2.6\sigma$ and $2.9\sigma$ for the $\Lambda$CDM and the $w_0$CDM models, respectively, where in the latter $w_0 = -0.985_{-0.024}^{+0.023}$.

\begin{table}[t]
    \centering
    \caption{Statistical significance of CPL compared to $\Lambda$CDM and $w_0$CDM in terms of the logarithm of the Bayes' factors $\ln \mathcal{B}_i^\mathrm{CPL}$ and number of standard deviations, $\#\sigma$. Positive values of $\ln \mathcal{B}_i^\mathrm{CPL}$ indicate evidence in favor of CPL.}%
    \label{tab:LCDM_w0CDM_CPL}
    \begin{ruledtabular}
    \begin{tabular}{lccccc}
        Statistic & Model & CBS & CB & CS & BS \vspace{.1em} \\ \hline\rule{0pt}{1.1em}%
        $\ln \mathcal{B}_i^\mathrm{CPL}$ & 	$\Lambda$CDM	 & 	$3.87$	 & 	$4.59$ & 	$-0.20$	 & 	$1.78$ 	 \\[0.25em]
        & 	$w_0$CDM	 & 	$4.06$	 & 	$4.41$ & 	$-0.21$	 & 	$-0.22$ 	 \\[1.em]
        $\#\sigma$ & $\Lambda$CDM & $2.55$ & $2.76$ & $0.43$ & $1.71$ \\[0.25em]
        & $w_0$CDM & $2.94$ & $3.04$ & $0.32$ & $0.46$ \\
    \end{tabular}
    \end{ruledtabular}
\end{table}

The reported tensions are in good agreement with the findings of~\cite{DESI:2025zgx}, ensuring that our pipelines allow to assess whether our new parametrizations are statistically favored over CPL, and to what extent these can remain closer to $w = -1$ in the past.

\begin{table*}[t]
    \centering
    \caption{Results for the best-fit parameters $(\Omega_\mathrm{m}, H_0, \Omega_\mathrm{b}, w_0, w_a)$, the asymptotic value of the EOS at infinitely large redshift $w_\infty$ (quantities are reported in italic since it is a secondary parameter), and the logarithm of the Bayes' factors $\ln \mathcal{B}_i^\mathrm{CPL}$. Positive values of the latter indicate evidence in favor of CPL.}%
    \label{tab:results}
    \begin{ruledtabular}
    \begin{tabular}{lcccccccc}
        Data & Param. & $\Omega_\mathrm{m}$ & $H_0$ & $\Omega_\mathrm{b}$ & $w_0$ & $w_a$ & $w_\infty$ & $\ln \mathcal{B}_i^\mathrm{CPL}$ \vspace{.1em} \\ \hline\rule{0pt}{1.1em}%
        CBS
        & 	CPL	 & 	$0.3108_{-0.0058}^{+0.0059}$	 & 	$67.58_{-0.61}^{+0.60}$	 & 	$0.04917_{-0.00090}^{+0.00093}$	& 	$-0.838_{-0.055}^{+0.055}$	 & 	$-0.60_{-0.22}^{+0.21}$	 & 	$\mathit{-1.43_{-0.17}^{+0.16}}$	 & 	-----------	 \\[0.5em]
        & 	Sqrt	 & 	$0.3105_{-0.0055}^{+0.0057}$	 & 	$67.59_{-0.60}^{+0.58}$	 & 	$0.04916_{-0.00089}^{+0.00092}$	 & 	$-0.832_{-0.057}^{+0.058}$	 & 	$-0.66_{-0.24}^{+0.23}$	 & 	$\mathit{-1.30_{-0.12}^{+0.12}}$	 & 	$0.333_{-0.036}^{+0.033}$	 \\[0.5em]
        & 	Sqrt$2$	 & 	$0.3107_{-0.0057}^{+0.0057}$	 & 	$67.59_{-0.59}^{+0.60}$	 & 	$0.04916_{-0.00090}^{+0.00091}$	 & 	$-0.839_{-0.056}^{+0.055}$	 & 	$-0.60_{-0.22}^{+0.21}$	 & 	$\mathit{-1.34_{-0.14}^{+0.13}}$	 & 	$0.113_{-0.035}^{+0.034}$	 \\[0.5em]
        & 	Sqrt$3$	 & 	$0.3108_{-0.0058}^{+0.0056}$	 & 	$67.58_{-0.59}^{+0.61}$	 & 	$0.04915_{-0.00089}^{+0.00092}$	 & 	$-0.839_{-0.055}^{+0.056}$	 & 	$-0.60_{-0.22}^{+0.21}$	 & 	$\mathit{-1.37_{-0.15}^{+0.14}}$	 & 	$0.074_{-0.030}^{+0.033}$	 \\[0.5em]
        & 	Sqrt$4$	 & 	$0.3108_{-0.0057}^{+0.0058}$	 & 	$67.58_{-0.60}^{+0.59}$	 & 	$0.04915_{-0.00090}^{+0.00091}$	 & 	$-0.839_{-0.055}^{+0.055}$	 & 	$-0.59_{-0.23}^{+0.20}$	 & 	$\mathit{-1.38_{-0.16}^{+0.14}}$	 & 	$0.049_{-0.033}^{+0.032}$	 \\[0.5em]
        & 	Tanh	 & 	$0.3107_{-0.0057}^{+0.0057}$	 & 	$67.57_{-0.60}^{+0.59}$	 & 	$0.04918_{-0.00089}^{+0.00089}$	 & 	$-0.833_{-0.057}^{+0.057}$	 & 	$-0.65_{-0.24}^{+0.22}$	 & 	$\mathit{-1.32_{-0.13}^{+0.12}}$	 & 	$0.190_{-0.032}^{+0.034}$	 \\[0.5em]
        & 	$2$Tanh	 & 	$0.3106_{-0.0057}^{+0.0058}$	 & 	$67.60_{-0.60}^{+0.60}$	 & 	$0.04914_{-0.00089}^{+0.00092}$	 & 	$-0.838_{-0.056}^{+0.056}$	 & 	$-0.61_{-0.23}^{+0.21}$	 & 	$\mathit{-1.40_{-0.16}^{+0.15}}$	 & 	$0.093_{-0.035}^{+0.033}$	 \\[0.5em]
        & 	$3$Tanh	 & 	$0.3108_{-0.0055}^{+0.0058}$	 & 	$67.57_{-0.60}^{+0.58}$	 & 	$0.04919_{-0.00089}^{+0.00091}$	 & 	$-0.838_{-0.054}^{+0.057}$	 & 	$-0.60_{-0.22}^{+0.21}$	 & 	$\mathit{-1.42_{-0.17}^{+0.16}}$	 & 	$0.089_{-0.034}^{+0.035}$	 \\[0.5em]
        & 	$4$Tanh	 & 	$0.3108_{-0.0057}^{+0.0057}$	 & 	$67.58_{-0.59}^{+0.60}$	 & 	$0.04915_{-0.00088}^{+0.00091}$	 & 	$-0.839_{-0.055}^{+0.057}$	 & 	$-0.60_{-0.22}^{+0.21}$	 & 	$\mathit{-1.42_{-0.17}^{+0.16}}$	 & 	$0.028_{-0.032}^{+0.032}$	 \\[1.em]
        CB
        & 	CPL	 & 	$0.354_{-0.020}^{+0.020}$	 & 	$63.5_{-1.7}^{+1.9}$	 & 	$0.0555_{-0.0030}^{+0.0030}$	 & 	$-0.40_{-0.20}^{+0.20}$	 & 	$-1.80_{-0.59}^{+0.59}$	 & 	$\mathit{-2.20_{-0.40}^{+0.38}}$	 & 	-----------	 \\[0.5em]
        & 	Sqrt	 & 	$0.364_{-0.024}^{+0.024}$	 & 	$62.6_{-1.9}^{+2.1}$	 & 	$0.0571_{-0.0036}^{+0.0036}$	& 	$-0.26_{-0.25}^{+0.24}$	 & 	$-2.37_{-0.78}^{+0.78}$	 & 	$\mathit{-1.94_{-0.31}^{+0.31}}$	 & 	$0.040_{-0.030}^{+0.029}$	 \\[0.5em]
        & 	Sqrt$2$	 & 	$0.357_{-0.023}^{+0.023}$	 & 	$63.2_{-1.9}^{+2.1}$	 & 	$0.0561_{-0.0034}^{+0.0034}$	& 	$-0.36_{-0.23}^{+0.23}$	 & 	$-1.94_{-0.68}^{+0.66}$	 & 	$\mathit{-1.99_{-0.35}^{+0.33}}$	 & 	$0.068_{-0.031}^{+0.030}$	 \\[0.5em]
        & 	Sqrt$3$	 & 	$0.356_{-0.022}^{+0.023}$	 & 	$63.3_{-1.8}^{+2.0}$	 & 	$0.0558_{-0.0033}^{+0.0034}$	& 	$-0.37_{-0.22}^{+0.22}$	 & 	$-1.87_{-0.67}^{+0.65}$	 & 	$\mathit{-2.04_{-0.37}^{+0.36}}$	 & 	$0.073_{-0.030}^{+0.030}$	 \\[0.5em]
        & 	Sqrt$4$	 & 	$0.356_{-0.022}^{+0.022}$	 & 	$63.3_{-1.8}^{+2.0}$	 & 	$0.0559_{-0.0033}^{+0.0032}$	& 	$-0.37_{-0.22}^{+0.22}$	 & 	$-1.87_{-0.64}^{+0.63}$	 & 	$\mathit{-2.09_{-0.38}^{+0.36}}$	 & 	$0.028_{-0.030}^{+0.030}$	 \\[0.5em]
        & 	Tanh	 & 	$0.360_{-0.024}^{+0.024}$	 & 	$63.0_{-1.9}^{+2.1}$	 & 	$0.0565_{-0.0035}^{+0.0036}$	& 	$-0.31_{-0.25}^{+0.24}$	 & 	$-2.17_{-0.76}^{+0.75}$	 & 	$\mathit{-1.96_{-0.35}^{+0.33}}$	 & 	$0.092_{-0.031}^{+0.031}$	 \\[0.5em]
        & 	$2$Tanh	 & 	$0.356_{-0.022}^{+0.023}$	 & 	$63.3_{-1.9}^{+2.0}$	 & 	$0.0558_{-0.0033}^{+0.0034}$	& 	$-0.37_{-0.22}^{+0.23}$	 & 	$-1.90_{-0.66}^{+0.65}$	 & 	$\mathit{-2.13_{-0.39}^{+0.38}}$	 & 	$0.036_{-0.032}^{+0.032}$	 \\[0.5em]
        & 	$3$Tanh	 & 	$0.354_{-0.022}^{+0.021}$	 & 	$63.5_{-1.7}^{+2.0}$	 & 	$0.0556_{-0.0034}^{+0.0031}$	& 	$-0.39_{-0.22}^{+0.21}$	 & 	$-1.82_{-0.61}^{+0.63}$	 & 	$\mathit{-2.15_{-0.39}^{+0.40}}$	 & 	$0.061_{-0.032}^{+0.032}$	 \\[0.5em]
        & 	$4$Tanh	 & 	$0.354_{-0.022}^{+0.021}$	 & 	$63.5_{-1.8}^{+2.0}$	 & 	$0.0555_{-0.0033}^{+0.0032}$	& 	$-0.40_{-0.22}^{+0.21}$	 & 	$-1.80_{-0.63}^{+0.63}$	 & 	$\mathit{-2.17_{-0.41}^{+0.41}}$	 & 	$0.055_{-0.031}^{+0.031}$	 \\[1.em]
        CS
        & 	CPL	 & 	$0.319_{-0.013}^{+0.014}$	 & 	$67.0_{-1.4}^{+1.3}$	 & 	$0.0498_{-0.0019}^{+0.0021}$	 & 	$-0.92_{-0.11}^{+0.11}$	 & 	$-0.29_{-0.56}^{+0.53}$	 & 	$\mathit{-1.20_{-0.46}^{+0.42}}$	 & 	-----------	\\[0.5em]
        & 	Sqrt	 & 	$0.321_{-0.012}^{+0.014}$	 & 	$66.8_{-1.4}^{+1.2}$	 & 	$0.0501_{-0.0018}^{+0.0021}$	& 	$-0.93_{-0.12}^{+0.11}$	 & 	$-0.25_{-0.61}^{+0.61}$	 & 	$\mathit{-1.10_{-0.32}^{+0.32}}$	 & 	$0.019_{-0.033}^{+0.032}$	 \\[0.5em]
        & 	Sqrt$2$	 & 	$0.319_{-0.012}^{+0.015}$	 & 	$67.0_{-1.4}^{+1.2}$	 & 	$0.0498_{-0.0018}^{+0.0022}$	& 	$-0.92_{-0.11}^{+0.11}$	 & 	$-0.28_{-0.58}^{+0.56}$	 & 	$\mathit{-1.15_{-0.39}^{+0.36}}$	 & 	$0.084_{-0.033}^{+0.033}$	 \\[0.5em]
        & 	Sqrt$3$	 & 	$0.319_{-0.012}^{+0.014}$	 & 	$67.0_{-1.4}^{+1.3}$	 & 	$0.0498_{-0.0018}^{+0.0022}$	& 	$-0.92_{-0.11}^{+0.11}$	 & 	$-0.28_{-0.56}^{+0.54}$	 & 	$\mathit{-1.17_{-0.40}^{+0.38}}$	 & 	$0.011_{-0.031}^{+0.035}$	 \\[0.5em]
        & 	Sqrt$4$	 & 	$0.319_{-0.012}^{+0.014}$	 & 	$67.1_{-1.4}^{+1.3}$	 & 	$0.0497_{-0.0018}^{+0.0021}$	& 	$-0.92_{-0.11}^{+0.11}$	 & 	$-0.30_{-0.55}^{+0.54}$	 & 	$\mathit{-1.19_{-0.41}^{+0.39}}$	 & 	$0.000_{-0.033}^{+0.034}$	 \\[0.5em]
        & 	Tanh	 & 	$0.320_{-0.013}^{+0.014}$	 & 	$66.9_{-1.3}^{+1.3}$	 & 	$0.0500_{-0.0018}^{+0.0020}$	& 	$-0.92_{-0.11}^{+0.11}$	 & 	$-0.27_{-0.58}^{+0.57}$	 & 	$\mathit{-1.13_{-0.34}^{+0.33}}$	 & 	$0.007_{-0.031}^{+0.034}$	 \\[0.5em]
        & 	$2$Tanh	 & 	$0.319_{-0.013}^{+0.014}$	 & 	$67.0_{-1.4}^{+1.3}$	 & 	$0.0498_{-0.0019}^{+0.0021}$	& 	$-0.91_{-0.11}^{+0.11}$	 & 	$-0.30_{-0.60}^{+0.56}$	 & 	$\mathit{-1.20_{-0.44}^{+0.41}}$	 & 	$0.042_{-0.031}^{+0.030}$	 \\[0.5em]
        & 	$3$Tanh	 & 	$0.319_{-0.013}^{+0.014}$	 & 	$67.0_{-1.4}^{+1.3}$	 & 	$0.0498_{-0.0020}^{+0.0022}$	& 	$-0.92_{-0.11}^{+0.12}$	 & 	$-0.28_{-0.61}^{+0.55}$	 & 	$\mathit{-1.20_{-0.48}^{+0.43}}$	 & 	$0.041_{-0.031}^{+0.035}$	 \\[0.5em]
        & 	$4$Tanh	 & 	$0.319_{-0.012}^{+0.015}$	 & 	$67.0_{-1.4}^{+1.3}$	 & 	$0.0498_{-0.0018}^{+0.0022}$	& 	$-0.92_{-0.11}^{+0.11}$	 & 	$-0.27_{-0.55}^{+0.56}$	 & 	$\mathit{-1.19_{-0.44}^{+0.43}}$	 & 	$0.026_{-0.031}^{+0.032}$	 \\[1.em]
        BS
        & 	CPL	 & 	$0.304_{-0.021}^{+0.015}$	 & 	$66.8_{-2.3}^{+1.7}$	 & 	$0.0496_{-0.0025}^{+0.0035}$	 & 	$-0.888_{-0.058}^{+0.060}$	 & 	$-0.19_{-0.43}^{+0.45}$	 & 	$\mathit{-1.08_{-0.38}^{+0.41}}$	 & 	-----------	 \\[0.5em]
        & 	Sqrt	 & 	$0.303_{-0.020}^{+0.015}$	 & 	$66.7_{-2.2}^{+1.7}$	 & 	$0.0498_{-0.0024}^{+0.0035}$	& 	$-0.887_{-0.063}^{+0.064}$	 & 	$-0.19_{-0.46}^{+0.50}$	 & 	$\mathit{-1.03_{-0.28}^{+0.31}}$	 & 	$0.037_{-0.030}^{+0.030}$	 \\[0.5em]
        & 	Sqrt$2$	 & 	$0.303_{-0.022}^{+0.015}$	 & 	$66.7_{-2.4}^{+1.8}$	 & 	$0.0497_{-0.0024}^{+0.0038}$	& 	$-0.888_{-0.057}^{+0.063}$	 & 	$-0.17_{-0.45}^{+0.49}$	 & 	$\mathit{-1.04_{-0.33}^{+0.37}}$	 & 	$0.102_{-0.031}^{+0.033}$	 \\[0.5em]
        & 	Sqrt$3$	 & 	$0.303_{-0.021}^{+0.015}$	 & 	$66.8_{-2.3}^{+1.7}$	 & 	$0.0497_{-0.0024}^{+0.0036}$	& 	$-0.889_{-0.057}^{+0.061}$	 & 	$-0.18_{-0.43}^{+0.45}$	 & 	$\mathit{-1.06_{-0.34}^{+0.37}}$	 & 	$0.040_{-0.030}^{+0.031}$	 \\[0.5em]
        & 	Sqrt$4$	 & 	$0.304_{-0.021}^{+0.015}$	 & 	$66.9_{-2.2}^{+1.7}$	 & 	$0.0496_{-0.0024}^{+0.0036}$	& 	$-0.889_{-0.057}^{+0.061}$	 & 	$-0.19_{-0.43}^{+0.46}$	 & 	$\mathit{-1.07_{-0.35}^{+0.38}}$	 & 	$0.025_{-0.028}^{+0.030}$	 \\[0.5em]
        & 	Tanh	 & 	$0.304_{-0.020}^{+0.014}$	 & 	$66.8_{-2.1}^{+1.7}$	 & 	$0.0496_{-0.0024}^{+0.0033}$	& 	$-0.889_{-0.060}^{+0.065}$	 & 	$-0.20_{-0.45}^{+0.47}$	 & 	$\mathit{-1.05_{-0.29}^{+0.31}}$	 & 	$0.042_{-0.031}^{+0.032}$	 \\[0.5em]
        & 	$2$Tanh	 & 	$0.304_{-0.021}^{+0.015}$	 & 	$66.8_{-2.3}^{+1.7}$	 & 	$0.0496_{-0.0024}^{+0.0036}$	& 	$-0.889_{-0.057}^{+0.062}$	 & 	$-0.20_{-0.45}^{+0.47}$	 & 	$\mathit{-1.07_{-0.36}^{+0.39}}$	 & 	$0.051_{-0.032}^{+0.034}$	 \\[0.5em]
        & 	$3$Tanh	 & 	$0.304_{-0.022}^{+0.015}$	 & 	$66.9_{-2.3}^{+1.7}$	 & 	$0.0496_{-0.0025}^{+0.0036}$	& 	$-0.887_{-0.058}^{+0.063}$	 & 	$-0.21_{-0.43}^{+0.48}$	 & 	$\mathit{-1.09_{-0.37}^{+0.42}}$	 & 	$0.039_{-0.031}^{+0.033}$	 \\[0.5em]
        & 	$4$Tanh	 & 	$0.304_{-0.022}^{+0.015}$	 & 	$66.8_{-2.3}^{+1.7}$	 & 	$0.0496_{-0.0025}^{+0.0035}$	& 	$-0.888_{-0.055}^{+0.063}$	 & 	$-0.19_{-0.46}^{+0.45}$	 & 	$\mathit{-1.08_{-0.40}^{+0.41}}$	 & 	$-0.006_{-0.032}^{+0.031}$	 \\
    \end{tabular}
    \end{ruledtabular}
\end{table*}

\subsection{Performance of the sigmoid families}

\subsubsection{Equation of state parameters}

First of all,  visual inspection of the values of $\ln \mathcal{B}_i^\mathrm{CPL}$ in Table~\ref{tab:results} reveals that almost all parametrizations perform equally well compared to CPL across all dataset combinations, except Sqrt and Tanh ($n = 1$), which are just slightly penalized over CPL for the CBS combination. This result is not unexpected since, as we advanced in Sec.~\ref{sec:parametrizations}, 
the latter parametrizations tend to diverge more strongly from the CPL model across a broader range of $a$, whereas as $n$ increases these families tend to remain closer to CPL. 

\begin{figure}[t]
    \centering
    \includegraphics{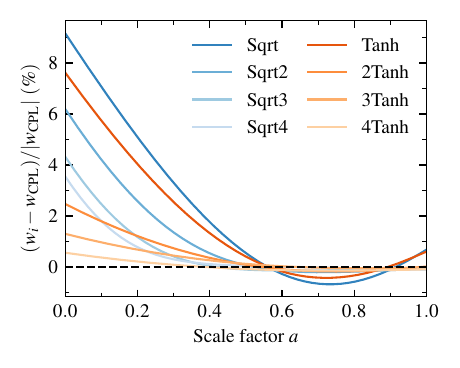}
    \vspace{-0.7em}
    \caption{The curves represent the relative difference of the DE EOS parameter between each proposed parametrization $w_i$ and CPL, defined as $(w_i - w_\mathrm{CPL})/\vert w_\mathrm{CPL} \vert$. Each curve corresponds to a specific family---Sqrt$n$ (blue) or $n$Tanh (orange)---and a representative value of $n$, indicated by the gradient of colors. The best-fit $(w_0, w_a)$ values from the CMB+BAO+SNeIa dataset combination are used for both $w_i$ and $w_\mathrm{CPL}$. Since all EOS parameters are negative, positive values on the vertical axis indicate a less phantom behavior compared to CPL. The black dashed line marks values of the EOS corresponding to $w_i = w_\mathrm{CPL}$. As $n$ increases, the deviation from CPL progressively decreases, illustrating the convergence of the proposed parametrizations towards the CPL limit.}%
    \label{fig:w_difference}
\end{figure}

As an example, let us consider the CBS combination. In this case, the points just raised are more clearly reflected in Fig.~\ref{fig:w_difference}, where we present the relative difference of the pressure-to-energy-density ratios of both families with respect to the CPL parametrization; positive values in the vertical axis indicate a less phantom behavior compared to CPL. We see that for $a \geq 0.5$ and values of $n \geq 2$ the families are completely indistinguishable from CPL, and those with $n = 1$ only deviate less than $1\%$. On the other hand, for $a < 0.5$, the relative differences are more significant, ranging from less than $1\%$ up to $9\%$. As the DESI and Pantheon+ data are mostly concentrated in the late-time universe ($z \lesssim 2$, which results in $a \gtrsim 1/3$), where DE is more strongly constrained, such data compilations offer strong enough constraints to identify these deviations, thus showing a slight preference for CPL compared to Sqrt and Tanh. With increasing values of $n$, the relative difference of $w(a)$ between the families and that of CPL is negligible in the local universe and much smaller at high redshifts, therefore performing equally well. This also means that deviations in the EOS of DE in the distant past cannot be probed by current surveys, opening the possibility for DE to display a strong nonlinear behavior (with a possible $\Lambda$ solution for $a \to 0$) in the framework of a more sophisticated model. Broadly, these conclusions also apply to the rest of data combinations explored in Table~\ref{tab:results}.

Now, let us turn our attention to the numerical results on the EOS parameters in Table~\ref{tab:results}. This is a good opportunity to highlight that the $w_0$ and $w_a$ parameters of the Sqrt$n$ and $n$Tanh are different parameters depending on the family and the value of $n$ since, properly speaking, these are parameters associated to different parametrizations. Even so, given the fact that these families tend to remain sufficiently close to CPL, it is justified to compare their values across the different parametrizations.

Following the trend aforementioned, an increase in the value of $n$ implies a better mimicry of the CPL parametrization, which is a result of the convergence of the parameters $w_0$ and $w_a$ to the values of $w_0^\mathrm{CPL}$ and $w_a^\mathrm{CPL}$, regardless of the data combination. In particular, from examination of their values in Table~\ref{tab:results} and Fig.~\ref{fig:w_difference}, it follows that the convergence is faster for the $n$Tanh parametrization, with $2$Tanh remaining closer to CPL than Sqrt$4$. Nevertheless, appreciable differences are found across the presented surveys.

On the one hand, the BS combination presents the largest values of $w_a$ (roughly $w_a \simeq -0.19 \pm 0.47$ for all parametrizations), being compatible with an EOS with no time evolution $(w_a = 0)$ by less than one standard deviation, and with current value of the EOS close to a cosmological constant, $w_0 \simeq -0.888 \pm 0.064$. This indicates that late-time data prefer values of the EOS that remain close to $\Lambda$, which is reinforced by the fact that the best-fit values of $w_\infty$ remain just slightly below the phantom crossing divide although compatible with $w(a) \geq -1$. In fact, this is the reason why the $w_0$CDM performs equally well compared to CPL and the remaining parametrizations; see Table~\ref{tab:LCDM_w0CDM_CPL}. According to the values reported in~\cite{Giare:2025pzu}, which also include SNeIa in the lower redshift range $z < 0.01$, the values of $w_a$ are marginally more negative though still compatible with no evolution.

On the other hand, for the CS combination, the values of $w_a$ are more negative than in the BS scenario with errors increasing accordingly, $w_a \simeq -0.28 \pm 0.58$, whereas the values of $w_0$ are closer to a cosmological constant with errors making it compatible with $\Lambda$ at less than $1\sigma$, $w_0 \simeq -0.92 \pm 0.11$. Recall from Table~\ref{tab:LCDM_w0CDM_CPL} that the CS combination exhibits the lowest deviation from $\Lambda$CDM.

The most statistically favored models compared to $\Lambda$CDM emerge from the CB combination, where the values of $w_0$ increase significantly, above $-0.40$, and those of $w_a$ decrease considerably, below $-1.8$, pointing towards a strong evolutionary behavior and a clear phantom crossing for all parametrizations. This is accompanied with the fact that a positive deceleration parameter ``best-fit'' is found for all parametrizations. In this sense, the absence of very low redshift data coming from SNeIa such as Pantheon+ seems to difficult an accordance with the currently accepted $q_0 < 0$ value. Interestingly, the Sqrt and Tanh parametrizations are almost not penalized over CPL regarding their Bayes factor values, and both exhibit a more violent time evolution but with higher values of the asymptotic EOS, $w_\infty \simeq - 1.95 \pm 0.33$ compared to that of CPL, $w_\infty^\mathrm{CPL} = -2.20_{-0.40}^{+0.38}$. This opens the possibility of considering parametrizations that deviate further from CPL at low redshift while keeping $w_\infty$ as close to $\Lambda$ as possible without strong statistical penalization.

If we now look at the trio CBS, the obtained values for $w_0$ and $w_a$ are a balance of all the previous cases discussed, with the current value of the EOS deviating from $\Lambda$, up to $w_0^\mathrm{Sqrt} = -0.832_{-0.057}^{+0.058}$, and with a clear time evolution but not as significant as in the CB combination, with a minimum value of $w_a^\mathrm{Sqrt} = -0.66_{-0.24}^{+0.23}$. We believe that these values of $w_0$ are mostly influenced by the inclusion of Pantheon+ supernovae, since this survey explores much lower redshifts than DESI (as low as $z = 0.01$ in our analysis), where DE seems to require values of the EOS close to $\Lambda$ in order to explain the data. Besides, in this case the Sqrt and Tanh parametrizations perform slightly worse than the remaining parametrizations, and therefore deviations from CPL at lower redshifts seem to be penalized more strongly.

\begin{figure}[t]
    \centering
    \includegraphics{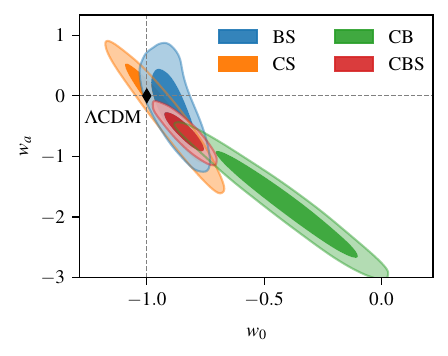}
    \vspace{-0.7em}
    \caption{Confidence-level regions of $68\%$ (shaded) and $95\%$ (light) of the CPL parametrization in the $(w_0, w_a)$ plane for various dataset combinations; CMB$+$BAO$+$SNeIa (CBS), CMB$+$BAO (CB), CMB$+$SNeIa (CS), and BAO+SNeIa (BS). All dataset combinations yield compatible results within $2\sigma$, even though the results are very sensitive to the redshift window of each combination. parametrizations other than CPL are not illustrated, as negligible shifts are anticipated for Sqrt$n$ and $n$Tanh according to the $(w_0, w_a)$ panel of Fig.~\ref{fig:triangle_FULL}.}%
    \label{fig:w0_wa_CPL}
\end{figure}

Based on the consistency with CPL of the parametrizations introduced in this work across all the data combinations, the obtained results confirm that the best-fit parameters are strongly sensitive to the redshifts explored by the data~\cite{Wolf:2025jlc, Wolf:2024eph, Wolf:2023uno, Shlivko:2024llw, Cortes:2024lgw}. The contours represented in the $(w_0, w_a)$ plane at confidence levels of $68\%$ and $95\%$ in Fig.~\ref{fig:w0_wa_CPL} for the CPL parametrization highlight this fact, with all the different dataset combinations being compatible at $1-2\sigma$. Furthermore, the redshift window explored by each late-time survey affects significantly the resulting values of the EOS parameters, but in any case all of these point towards values close to $\Lambda$ below $z = 1$. This, in turn, is strongly related to the pivot redshift, i.e., the value of the redshift at which $w(a)$ is more tightly constrained~\cite{Huterer:2000mj, Albrecht:2006um, Linder:2006xb}.

The statistical significance of the results of the changes is so small that this is rather an indication of the robustness of the CPL behavior. From here, we can extract two important conclusions. 

\begin{figure*}[t]
    \centering
    \includegraphics{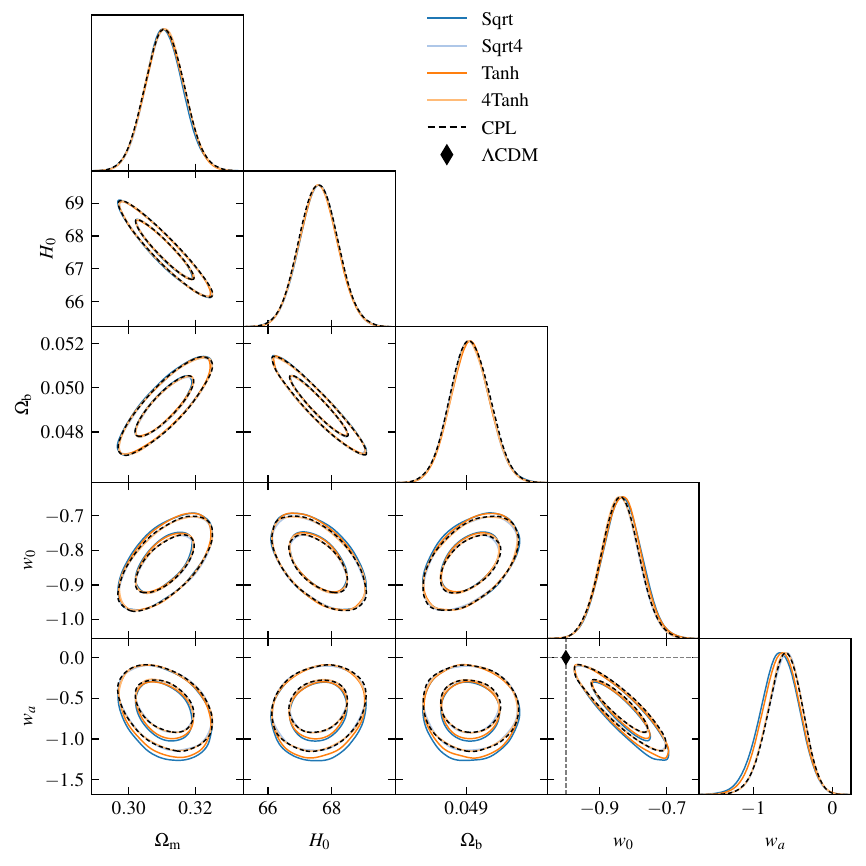}
    \vspace{-0.7em}
    \caption{Marginalized posterior distributions of $(\Omega_\mathrm{m}, \Omega_\mathrm{b}, H_0, w_0, w_a)$ derived from the CMB$+$BAO$+$SNeIa combination for the CPL parametrization, along with two exemplars each from the Sqrt$n$ and $n$Tanh families. The $\Lambda$CDM value of the EOS is included in the $(w_0, w_a)$ plane as a reference.}
    \label{fig:triangle_FULL}
\end{figure*}

On the one hand, phantom crossing is a statistically favored feature when the CMB and BAO data are combined~\cite{Linder:2007ka,DESI:2025zgx}, as the parametrizations that deviate further from CPL at small redshifts perform (just slightly) worse than CPL. To illustrate this fact, we previously represented the EOS parameter of CPL, Sqrt and Tanh for the CBS combination in Fig.~\ref{fig:w_plot}. Clearly, all of them cross the phantom divide line $w = -1$ at very low redshifts, and since parametrizations with larger values of $n$ remain closer to CPL and perform comparably, this crossing occurs for all the parametrizations. The values reported in Table~\ref{tab:results} reinforce this claim, as $w_\infty < -1$ and $w_0 > -1$ for all the dataset combinations (with varying statistical significances). The reason behind the need of CMB and BAO is twofold; DESI enhances the significance of evolving DE, and CMB requires the crossing of the phantom divide line in order to explain the measurement of the shift parameters $l_a$ and $R$ \cite{DESI:2025zgx, Linder:2007ka}. 

On the other hand, further hints for the phantom nature in the past cannot be inferred by current cosmological data, since larger differences in the early-time values of $w(a)$ are neither favored nor disfavored. This means that models (presumably arising from fundamental actions) that perform strongly nonlinearly could reduce the phantomness of DE in the past while turning back to a CPL-like EOS at late times.

\subsubsection{Matter, baryons, and Hubble parameter}

Equipped with the results of the pressure-to-energy-density ratio parameters $(w_0, w_a)$, we direct our focus towards the remaining cosmological parameters: $\Omega_\mathrm{m}$, $H_0$ and $\Omega_\mathrm{b}$. The corner plot in Fig.~\ref{fig:triangle_FULL} presents the marginalized posteriors of $(\Omega_\mathrm{m}, H_0, \Omega_\mathrm{b}, w_0, w_a)$ for the CBS combination, comparing the CPL parametrization with the key representatives of Sqrt$n$ and $n$Tanh. A visual inspection reveals that posteriors involving $(\Omega_\mathrm{m}, H_0, \Omega_\mathrm{b})$ are visually indistinguishable from each other, while $w_0$ contours for Sqrt and Tanh (fourth row in Fig.~\ref{fig:triangle_FULL}) exhibit slight shifts, and $w_a$ contours (fifth row in Fig.~\ref{fig:triangle_FULL}) shift more appreciably towards negative values.

Analysis of Table~\ref{tab:results} indicates minimal differences across parametrizations within a fixed data framework, except for Sqrt and Tanh ($n = 1$) in the CMB$+$BAO setting, showing deviations up to $\sim 0.5\sigma$. This occurs because when parametrizing the DE EOS the Hubble function is the primary cosmological observable influenced, being dependent on the integrated energy density of DE, effectively smoothing out variations on $w(a)$. Consequently, large variations in $w(a)$ lead to small shifts in the Hubble function, and since many observables involve further integration of the latter, the differences at the level of the EOS are even more attenuated~\cite{Maor:2000jy, dePutter:2008wt}. For this reason, $w(a)$ parameters are less tightly constrained than other cosmological parameters~\cite{Maor:2001ku}.

Nevertheless, as occurred with the EOS parameters, significant discrepancies can be found when comparing results from different survey combinations. The largest of these are found for the CB combination, which exhibit large values of the fractional matter and baryon densities, $\Omega_\mathrm{m}^\mathrm{CPL} = 0.354 \pm 0.020$ and $\Omega_\mathrm{b}^\mathrm{CPL} = 0.0555 \pm 0.0030$, and much lower values of the Hubble constant $H_0 = 63.5_{-1.7}^{+1.9}$ compared to the remaining pairs and CBS. The extremely low value of the Hubble parameter can also be related to the large values obtained for $w_0$, as already noted in~\cite{Colgain:2025nzf} and references therein. Presumably, this is due to the lack of SNeIa data, which tends to force lower values of the matter density and larger values of the Hubble constant (see Table~\ref{tab:results}). The reason behind the large values of $\Omega_\mathrm{b}$ is that our CMB data compression imposes a prior over the reduced fractional baryon density $\Omega_\mathrm{b} h^2$. Consequently, reducing the value of the Hubble constant requires an increase in the fractional baryon density in order to fit this data point.

\section{Summary and conclusions}\label{sec:conclusions}

The recent measurements of DESI BAO~\cite{DESI:2024mwx,DESI:2025zgx} challenged our understanding of the universe, pointing out that the long standing $\Lambda$ might not be the last word. When considering a dynamic DE component through the lens of the CPL parametrization, the data combinations of CMB$+$DESI and CMB$+$DESI$+$Pantheon+ indicate statistical significances of $2.8\sigma$ and $2.6\sigma$, respectively, supporting evolving DE. Specifically, the findings highlight the region where $w_0 > -1$ and $w_a < 0$ (Fig.~\ref{fig:w0_wa_CPL}), suggesting that DE displays present quintessence features with a past phantomlike dynamics.

Despite the previous findings, there is still no solid theoretical basis for the unusually extreme negativity of the pressure contribution coming from DE and the recent crossing of the phantom divide. This prompts the inquiry into the observational preference for phantom crossing and the degree to which DE diverged from a cosmological constant in the past ($w = -1$). In order to tackle these issues, we explored DE parametrizations which perform linearly at recent times and at the same time allow for the possibility of performing nonlinearly in the distant past, hoping to reduce the phantomness exhibited by CPL. We found a simple route appealing to suitably chosen sigmoid functions, and in particular we proposed two families of parametrizations, namely Sqrt$n$ and $n$Tanh. Both of these share the feature that for increasing values of $n$ the mimicry with CPL extends to a larger redshift range, making it challenging for current surveys to detect discrepancies in the redshift range where late-time data rely. To asses the statistical significance of the proposed models with respect to CPL, we considered three main datasets; DESI BAO~\cite{DESI:2025zgx}, the Pantheon+ SNeIa survey~\cite{Scolnic:2021amr, Peterson:2021hel, Carr:2021lcj, Brout:2022vxf}, and the compressed CMB shift parameters~\cite{Bansal:2025ipo, Planck:2019nip, ACT:2023kun}. The results and statistical significances were obtained using MCMC and nested sampling pipelines, and the main results are presented in Tables~\ref{tab:LCDM_w0CDM_CPL} and~\ref{tab:results}. As a consistency check, we compared the performance of the CPL with both the $\Lambda$CDM and $w_0$CDM in Table~\ref{tab:LCDM_w0CDM_CPL}, finding that the reported statistical significances agree with those presented by the DESI Collaboration~\cite{DESI:2025zgx}. 

Regarding our parametrizations and attending to their Bayes' factors, we find no statistical preference compared to CPL for any of the survey combinations according to Jeffreys' scale, pointing towards the robustness of the linear parametrization. Indeed, both families rapidly converge to CPL, becoming nearly indistinguishable for $n = 4$. The CMB$+$SNeIa and DESI$+$SNeIa combinations exhibit little evolution with values of $w_a$ close to zero and asymptotic high-redshift values $w_\infty$ compatible with a cosmological constant. The picture changes significantly when considering the CMB$+$BAO and CMB$+$BAO$+$SNeIa combinations, where DE presents a more dramatic time evolution and a strongly phantom behavior in the distant past. The inclusion of both DESI and CMB reinforces the necessity of phantom crossing; on the one hand, DESI points towards evolutionary DE, on the other hand, the measurement of the shift parameters $l_a$ and $R$ provided by the CMB requires crossing of the phantom divide line delimited by $w_\Lambda = -1$~\cite{DESI:2025zgx,Linder:2007ka}. Conversely, Fig.~\ref{fig:w0_wa_CPL} shows that SNeIa, owing to their large statistical weight, tighten the early-time related CMB$+$BAO constraints and shift the $(w_0, w_a)$ contours towards the top-left region, closer to $\Lambda$CDM. Clearly, the results of the parametrization of the EOS depend crucially on the election of the dataset~\cite{Wolf:2025jlc, Wolf:2024eph, Wolf:2023uno, Shlivko:2024llw, Cortes:2024lgw}; see also Fig.~\ref{fig:w0_wa_CPL}.

Turning to the remaining cosmological parameters $(\Omega_\mathrm{m}, H_0, \Omega_\mathrm{b})$, our parametrizations yield equal values compared to those of CPL. The double integration of the EOS required to compute the cosmological distances, which at the end of the day are the observables required for cosmological inference, smooths out any deviation at the level of the pressure-to-energy-density ratio. The largest discrepancies between datasets are found for the CMB$+$DESI combination, where large values of the fractional matter and baryon energy densities $\Omega_\mathrm{m}$ and $\Omega_\mathrm{b}$ and low values of the Hubble constant $H_0$ are obtained compared to the remaining combinations. In contrast, SNeIa data from the Pantheon+ sample generally indicates reduced $\Omega_\mathrm{m}$ and increased $H_0$ values, with the latter implying decreased $\Omega_\mathrm{b}$ values due to the constraints on $\Omega_\mathrm{b} h^2$ coming from BBN or CMB.

In conclusion, the CPL parametrization remains a robust phenomenological EOS which captures the dynamics of DE in the redshift window explored by current surveys; this suggests that the crossing of the phantom divide line is a feature of DE. This trend has been further reinforced by several independent analyses~\cite{Ozulker:2025ehg, Scherer:2025esj} based on different SNeIa samples, including Union3~\cite{Rubin:2023jdq} and the Year~5 data release of the Dark Energy Survey~\cite{DES:2024jxu}, as well as by nonparametric reconstructions of the EOS such as the weighted function regression method~\cite{Gonzalez-Fuentes:2025lei}. However, since the early-epoch EOS values appear irrelevant as long as matter domination persists, there is still room for models exhibiting complex, nonlinear behavior---such as nonminimally coupled scalar fields~\cite{Wolf:2024stt,Wolf:2024eph} or multifield scenarios in which none of the individual fields is inherently phantom but their combined effective EOS displays a phantom divide crossing~\cite{Gomez-Valent:2025mfl}---capable of reproducing $\Lambda$ throughout much of the cosmic history.

\acknowledgments
M.~A.~acknowledges support from the Basque Government Grant No.~PRE\_2024\_1\_0229. M.~A., I.~A., and R.~L.~are supported by the Basque Government Grant No.~IT1628-22, and by Grant No.~PID2021-123226NB-I00 (funded by MCIN/AEI/10.13039/501100011033, by ``ERDF A way of making Europe''). I.~A.~is also supported by the Grant Juan de la Cierva funded by MICIU/AEI/10.13039/501100011033 and by “ESF+”. The research of V.~S.~is funded by the Polish National Science Centre Grant No.~DEC-2021/43/O/ST9/00664. This article is based upon work from the COST Action CA21136 - “Addressing observational tensions in cosmology with systematics and fundamental physics (CosmoVerse)”, supported by COST - “European Cooperation in Science and Technology”.

\section*{Data Availability}
No data were created or analyzed in this study.

\bibliographystyle{apsrev4-2-titles}
\bibliography{bib}
\end{document}